\documentclass[12pt,preprint,flushrt]{aastex}
\usepackage[table]{xcolor}
\usepackage[normalem]{ulem}
\begin{document}

\title{First Search for an X-ray -- Optical Reverberation Signal in an
  Ultraluminous X-ray Source}
\author{Dheeraj R.~Pasham$^{1,2}$, Tod E. Strohmayer$^{1,2}$, S.~Bradley Cenko$^{1,2,3}$,
Margaret L. Trippe$^{4}$, Richard F. Mushotzky$^{2,3}$, Poshak Gandhi$^{5}$
}
\altaffiltext{1}{Code 661, Astrophysics
Science Division, NASA's Goddard Space Flight Center, Greenbelt, MD
20771, USA}
\altaffiltext{2}{Joint Space-Science Institute, University of Maryland, College
Park, MD 20742, USA}
\altaffiltext{3}{Department of Astronomy, University of Maryland, College Park, MD 20742, USA}
\altaffiltext{4}{Johns Hopkins University Applied Physics Laboratory, Laurel, MD, USA}
\altaffiltext{5}{School of Physics \& Astronomy, University of Southampton, Southampton, SO17 1BJ, U.K.}
\email{Email: dheerajrangareddy.pasham@nasa.gov}

\begin{abstract}
Using simultaneous optical (VLT/FORS2) and X-ray ({\it XMM-Newton}) 
data of NGC 5408, we present the first ever attempt to search for a 
reverberation signal in an ultraluminous X-ray source (NGC 5408 X-1). 
The idea is similar to AGN broad line reverberation mapping where a 
lag measurement between the X-ray and the optical flux combined with 
a Keplerian velocity estimate should enable us to weigh the central 
compact object. We find that although NGC 5408 X-1's X-rays are 
variable on a timescale of a few hundred seconds (RMS of 
9.0$\pm$0.5\%), the optical emission does not show any statistically 
significant variations. We set a 3$\sigma$ upper limit on the RMS 
optical variability of 3.3\%. The ratio of the X-ray to the optical 
variability is an indicator of X-ray reprocessing efficiency. In 
X-ray binaries, this ratio is roughly 5. Assuming a similar ratio 
for NGC 5408 X-1, the expected RMS optical variability is 
$\approx$2\% which is still a factor of roughly two lower than what was 
possible with the VLT observations in this study. We find marginal 
evidence (3$\sigma$) for optical variability on a $\sim$ 24 hour 
timescale. Our results demonstrate that such measurements can be made, 
but photometric conditions, low sky background levels and longer 
simultaneous observations will be required to reach the optical 
variability levels similar to X-ray binaries.


\end{abstract}


\vfill\eject

\newpage

\section{Introduction}
X-ray bright, off-nuclear point sources in nearby galaxies whose
isotropic luminosities exceed the Eddington value of a 25 $M_{\odot}$
black hole, i.e., $\ga$ 3$\times$10$^{39}$ ergs
s$\textsuperscript{-1}$, are referred to as ultraluminous X-ray
sources (ULXs).  The main debate concerning these objects is whether
they are stellar-mass black holes (mass range of 3-25 $M_{\odot}$)
circumventing the Eddington limit via high accretion and/or emission
(e.g., King et al. 2001; K\"ording et al. 2002; Begelman et al. 2002;
Poutanen et al. 2007; Gladstone et al. 2009), or if they are powered
by intermediate-mass black holes (IMBHs: mass range of a few 100-1000
$M_{\odot}$) accreting below the Eddington limit (e.g., Colbert \&
Mushotzky 1999; Miller et al. 2004, 2013). Recent studies suggest that
ULXs are very likely an inhomogeneous sample of both stellar and
intermediate-mass black holes (e.g., Farrell et al. 2009; Motch et
al. 2014; Pasham et al. 2014, 2015; Mezcua et al. 2015); and in some
rare cases may even be powered by neutron stars (Bachetti et
al. 2014).

A straightforward way to solve the ULX mass problem is by measuring
the dynamical masses of their compact objects. This involves
identifying the compact object's companion star (an optical
counterpart) and Doppler tracking its radial velocity (using
spectroscopic lines) from which one can extract the orbital
period of the system and the companion's radial velocity
semi-amplitude. These two quantities can be combined to construct the
so-called mass function, which would enable a lower limit on the black
hole mass (see, for example, Soria et al. 1998 for relevant
equations). Although very challenging, there have been a handful of
such attempts that have succeeded in weighing stellar-mass black holes
in ULXs (e.g., Liu et al. 2013; Motch et al. 2014; See, however,
Roberts et al. 2011, Cseh et al. 2013, where the mass could not be
tightly constrained). This method is difficult primarily because ULX
optical counterparts are extragalactic, and hence faint with V-band
magnitudes in the range of 22-24 (e.g., Tao et al. 2011; Gladstone et
al. 2013).  Extracting their optical spectra is not only challenging
but also very expensive--requiring an 8-m class telescope (e.g., Cseh
et al. 2013).

In the absence of dynamical mass constraints various indirect
methods--that have been well-calibrated against stellar-mass black
holes--have been used. For example, it is known in stellar-mass black
hole systems that at a given X-ray energy spectral power-law index,
the centroid frequency of the low-frequency quasi-periodic oscillation
(QPO) scales inversely with the black hole mass (e.g., Sobczak et
al. 2000; Vignarca et al. 2003; Shaposhnikov \& Titarchuk
2009). Assuming that the ULX mHz QPOs are analogous to the
low-frequency QPOs of stellar-mass black holes (0.2-15 Hz: Casella et
al. 2005), various authors have estimated some ULX black hole masses
to be in the range of a few$\times$(100-1000) $M_{\odot}$ (e.g.,
Dewangan et al.  2006; Strohmayer \& Mushotzky 2009; Feng et al. 2010;
Rao et al. 2010; Dheeraj \& Strohmayer 2012). However, this underlying
assumption of mHz QPOs being the analogs of the low-frequency QPOs has
been questioned by some authors (Middleton et al. 2011). In a few
cases, in addition to the mHz QPOs, 3:2 frequency ratio high-frequency
QPO analogs have also been discovered. Such twin pairs not only confirm 
that the mHz QPOs in these systems are indeed the analogs of the stellar-mass
black hole low-frequency QPOs, but also enable independent, accurate
black hole masses. Assuming the inverse mass--to--QPO frequency
relation of stellar-mass high-frequency QPOs, the detected 3:2 pairs
in M82 X-1 and NGC 1313 X-1 imply a black hole mass of 428$\pm$105
$M_{\odot}$ (Pasham et al. 2014), and 5000$\pm$1300 $M_{\odot}$
(Pasham et al.  2015), respectively.

Another direct method that can be employed to weigh black holes--but
has not been hitherto exploited for ULXs--is the so-called
reverberation mapping technique. It is now known that a significant
fraction of the optical emission from accreting X-ray binaries is due
to X-ray reprocessing in the surrounding accretion disk (van Paradijs
\& McClintock 1994; Reynolds \& Miller 2013), and current evidence
suggests that this may be the case in some ULXs (Tao et al. 2011;
Gladstone et al. 2013). In simple models of this process the optical
emission results from X-ray irradiation of the outer portions of the
disk by the central continuum. The optical emission is thus correlated
with the X-ray emission but delayed by the light travel time effects
(Hynes et al. 1998; Gandhi et al. 2010). Measurement of the time
delay, $\tau$, will provide a direct estimate of the accretion disk's
(and hence the binary's) size ($R \approx c\tau$, where $c$ is the
speed of light). A mass estimate can follow in a manner analogous to
AGN reverberation mapping (Peterson \& Wandel 2000). The difference is
that in ULXs the optical flux is expected to reverberate in response
to the variations in the central X-ray flux while in AGN the optical
broad line flux responds to changes in the optical
continuum. Nevertheless, a lag measurement gives a size scale of the
accretion disk which can then be combined with measured line widths
($\Delta$$v$) due to the Keplerian motions in the disk. The mass of
the ULX's compact object ($M$) can then be estimated as,
\begin{equation}
	M = \frac{f c \tau \Delta v^{2}}{G}
\end{equation}
where $G$ is the Gravitational constant and $f$ is a geometric
correction factor whose value is expected to be of order unity.

In practice, however, the reprocessed optical emission can originate from a wide 
range of accretion disk radii and from even further out from the face 
of the companion star illuminated by the X-rays (e.g., O'Brien et al. 
2002). Therefore, in order to accurately measure the black hole's mass 
one would need to particularly measure the lag between the X-rays and 
the reprocessed optical flux originating from the part of the accretion 
disk that is spatially coincident with the optical emission lines. This 
can, in principle, be achieved by simultaneous X-ray and narrow band 
optical photometry (e.g., see Mu{\~n}oz-Darias et al. 2007 for the case 
of Sco X-1). But even with simultaneous X-ray and broad band optical 
observations it should be possible to clearly distinguish a ULX IMBH 
from a ULX stellar-mass black hole. This is simply because for a fixed 
binary orbital period a system hosting an IMBH (say, 1000 $M_{\odot}$) 
will be significantly larger than one with a stellar-mass black hole 
primary (say, 10$M_{\odot}$). For example, the binary separation scales 
as $M_{tot}^{1/3}$, where $M_{tot}$ is the total system mass and 
additionally, the radius of the Roche lobe of the accretor, which sets 
an upper limit on the size of the accretion disk, also grows significantly 
with $M_{tot}$ (e.g., Eggleton 1983). On the other hand, the situation can be
more complicated as the optical emission from X-ray binaries is not 
always from just X-ray reprocessing. Details of non-reprocessing optical 
emission and references thereof are discussed briefly in section 6. The
proposed reverberation methodology will work when the reprocessed optical 
emission component can be confidently identified. 

NGC 5408 X-1 (hereafter, X-1) is a ULX with an average X-ray (0.3-10
keV) luminosity of $\approx$ 10$^{40}$ ergs s$\textsuperscript{-1}$
(see, Table 4 of Dheeraj \& Strohmayer 2012). Based on its high
luminosity and the observed mHz QPO frequency range of 10-40 mHz
(Dheeraj \& Strohmayer 2012; Caballero-Garc{\'{\i}}a et al. 2013), it
has been suggested to host an IMBH with mass anywhere between a
few$\times$(100-1000) $M_{\odot}$ (Dheeraj \& Strohmayer
2012). Gris{\'e} et al. (2012) modeled the X-ray/UV/optical/NIR
spectral energy distribution (SED) of the ULX and concluded that its
optical emission is consistent with originating from an X-ray
irradiated accretion disk. In addition, deep optical spectroscopy of
this source has revealed the presence of a He II $\lambda$4686 $\textrm
{\AA}$ emission line whose broad component has an average full width
at half maximum (FWHM) of $\approx$ 780$\pm$64 km
s$\textsuperscript{-1}$ (Cseh et al. 2011; 2013). Based on the
observed $\approx$ 13\% variability of this broad emission component
over a time span of roughly 4 years and its Gaussian profile, Cseh et
al. (2011, 2013) argued that this broad line arises from Keplerian
motion within an accretion disk rather than a donor star. Assuming
that to be the case, if one can measure the distance of this optical
broad line emitting site from the central black hole, it can be
combined with the line width to directly measure the mass of the black
hole using equation (1). It can be noted straightforwardly from this 
equation that for X-1's black hole mass in the range of a 
few$\times$(100-1000) $M_{\odot}$ the expected time lags based on the 
line width are a few$\times$(100-1000) s. The maximum of the optical 
and the X-ray time resolution sets a lower limit on the time lag, and 
hence the black hole mass that can be probed with this method. In this 
proof-of-concept study for X-ray -- optical reverberation, we are 
sensitive to black hole masses greater than 600 $M_{\odot}$ because of an optical
time resolution of 440 s.






The article is arranged as follows. In section 2 we describe the
optical data acquired by the Very Large Telescope (VLT) and the X-ray
data taken with {\it XMM-Newton}.  In section 3, we describe the
procedure we used to extract the optical light curve while in section
4 we describe the X-ray data analysis. We present the cross-correlation
analysis in section 5 and discuss the implications and future prospects 
in section 6.

\section{Data Description}
The FOcal Reducer Spectrograph 2 (FORS2) on the VLT (UT1, ``Antu'')
observed the field of view containing X-1 ($\alpha = 14^{\rm h}03^{\rm
  m}19.62^{\rm s}$ and $\delta = -41^{\circ}22'58\farcs54$; Lang et
al. 2007) in imaging mode on three consecutive nights--for
approximately 16 ks each night--starting from 2014 February 12 until
2014 February 14$\footnote{Additional data was obtained on Feb
  11$\textsuperscript{th}$ and the 12$\textsuperscript{th}$ in the
  so-called HIT--High Time Resolution--mode, which was not useful for
  this analysis due to lack of reference stars for calibration in the
  smaller field of view.}$. The typical exposure times were 165 s on 
the night of the 12$\textsuperscript{th}$ and 120 s on the nights of  
the 13$\textsuperscript{th}$ and the 14$\textsuperscript{th}$. The 
source was also observed with {\it XMM-Newton} for approximately 
2$\times$36 ks on the nights of the 11$\textsuperscript{th}$ and the 
13$\textsuperscript{th}$ providing roughly 16 ks of useful simultaneous 
X-ray and optical photometric data. As discussed earlier, the optical
exposure time sets an absolute lower limit on the time lag, and hence 
the black hole mass that can be probed. As the time resolution of 120 
s is much longer than the typical X-ray -- optical lags of a few tens of 
seconds in stellar-mass X-ray binaries, these measurements are only 
sensitive to weigh IMBHs with masses greater than a few 100 $M_{\odot}$. A summary of the 
observational setup of the imaging mode is outlined in Table 1.

Since our main goal is to search for optical reverberation, we chose a
standard FORS2 g\_HIGH filter with an effective wavelength of 4700
$\textrm {\AA}$ and a FWHM of 1150 $\textrm {\AA}$, which is the band
that provided the most optical signal (see Figure 4 of Kaaret \& Corbel 
2009). Although the He II $\lambda$4686 $\textrm {\AA}$ broad emission 
line falls in the same band it is expected to have negligible 
contribution to the overall optical variability. The He II $\lambda$4686 
$\textrm {\AA}$ flux is typically a factor of $\approx$ 80 lower than the 
optical continuum flux (see Table 3 of Kaaret \& Corbel 2009). The same filter 
was used on all the nights.

\section{Analysis: Optical Photometry}
FORS2 is equipped with a mosaic of two 2k$\times$4k MIT CCDs (CHIP1 \&
CHIP2) with a pixel size of 15$\times$15 $\mu$m. Although the
intrinsic pixel scale is 0.126$^{\prime\prime}$ the standard data
binning of 2$\times$2 was used during all the observations providing a
final pixel scale of 0.252$^{\prime\prime}$. The observations were
carried out such that the source was always on CHIP1.  We used the
Image Reduction and Analysis Facility (IRAF)$\footnote{IRAF is
  distributed by the National Optical Astronomy Observatory, which is
  operated by the Association for Research in Astronomy, Inc., under
  cooperative agreement with the National Science Foundation.}$
software for calibration and other image analysis.

\subsection{Imaging Mode Data Analysis}
We performed the following analysis to generate X-1's optical light curves:

1) Basic data reduction: We first reduced the data through bias
subtraction and flat fielding followed by the trimming of the vignetted
regions of the images.

2) Data screening: X-1's optical counterpart is faint with a V-band
magnitude of 22.4--as measured in the F547M filter of HST (Gris{\'e}
et al. 2012). We found that including images with seeing worse than
1$^{\prime\prime}$ resulted in unusually large error bars on the light
curve measurements compared with the expected error bars from Poisson statistics. 
We therefore removed images with seeing worse
than this value from our analysis.

3) Image alignment: We aligned the images to a relative precision of
$\la$ 10\% of a pixel using the {\it Astromatic} software tools {\tt
  SExtractor} (Bertin \& Arnouts 1996), {\tt SCAMP} (Bertin 2006), and
{\tt SWarp} (Bertin et al. 2002).

4) Image subtraction: We performed image subtraction using {\tt
  HOTPANTS}$\footnote{http://www.astro.washington.edu/users/becker/v2.0/hotpants.html}$,
which is a custom modification of the ISIS algorithm developed by
Alard \& Lupton (1998).  We validated the subtracted images by
randomly selecting 1000 locations around X-1 and ensuring their light
curves were all consistent with being constant.

5) Extract X-1's optical light curve: Finally, we extracted X-1's
light curve by performing aperture photometry on the subtracted
images.

We now discuss in detail each of the above steps.

\subsubsection{Image Reduction:}
We first reduced all the images following the standard procedure of
subtracting the bias followed by division by a bias subtracted,
normalized flat field. We normalized the bias subtracted flat field by
its mode value. We used the IRAF tasks {\tt zerocombine} and {\tt
  flatcombine} to construct the median bias and the flat field images,
respectively. After the initial calibration, we trimmed the vignetted
regions of the images.

\subsubsection{Data Screening:}
On the night of the 13$\textsuperscript{th}$ the seeing (Point Spread
Function's (PSF's) FWHM) varied between 
0.5$^{\prime\prime}$-1.6$^{\prime\prime}$ with the seeing degrading
towards the end of the night. After the screening criterion of 
excluding images with seeing worse than 1$^{\prime\prime}$, we had 73 
images of roughly 120 s exposure each. This gave us a total temporal 
baseline of 73$\times$(120+27) s $\approx$ 10.7 ks (here 27 s refers 
to the image readout time). The airmass was below 2 all throughout 
the night. A sample image with the entire field of view (CHIP 1) is 
shown in Figure 1.

\subsubsection{Image Alignment:}
We first obtained the plate solution--using the IRAF task {\tt
  ccmap}--of an image taken halfway through the night of the
13$\textsuperscript{th}$. While the seeing worsened from the beginning 
to the end of the night, the airmass and the overall sky background
(due to the moon) improved in a more or less monotonic fashion.
Therefore, the optimum image for extracting the sources was the one at
the middle of the image stack. Using 20 pointlike sources in the 2MASS
catalog, we were able to achieve an absolute astrometric accuracy of
$\approx$ 0.1$^{\prime\prime}$ ($\approx$ 0.07$^{\prime\prime}$ in
both the RA and the declination coordinates).

We then noticed that even the images taken on the same night were slightly
misaligned with respect to each other by a fraction of a pixel to a few 
pixels, despite the fact that the telescope was not dithered between 
successive exposures. We corrected this as follows:

(a) First, we extracted--using {\tt SExtractor}--the locations of
various point sources in the reference image above with absolute astrometry. 

(b) We then used {\tt SCAMP} to obtain an astrometric solution for
each image to match with the positions of point sources derived from
the reference image in the above step. We only allowed for linear
distortions, i.e., the {\tt SCAMP} parameter {\tt DISTORT\_DEGREES}
was set to 1.

(c) Finally, we used {\tt SWarp} to re-sample each image on a
1672$\times$955 pixel grid as per the astrometric solution derived in
the above step. In essence, we aligned every image with respect to the
image in step (a).

The choice of the image in step (a) is not critical.  We repeated the
entire analysis with other images to find that the results are the
same.

We tested the accuracy of the image alignment by tracking the centroid
positions--($x$,$y$)--of a sample of stars in the aligned images. We
used {\tt SExtractor} to estimate the centers of 20 point sources in
all the 73 images after alignment. For the sake of convenience, we used
the same point sources used for plate solution above. These stars are 
indicated by blue colored circles and referenced as A followed by a number in Figure
1. Table 2 describes their WCS (J2000) coordinates and the RMS
variation in units of fraction of a pixel after alignment. The
observed RMS variation of 3-12\% suggests that the images are aligned
to less than one-eighth of a pixel with respect to each other.

\subsubsection{Image Subtraction:}
After aligning the images, we combined three consecutive
images to improve the signal-to-noise.  We then employed {\tt
  HOTPANTS} to carry out image subtraction on these 24 (73 over 3)
images. We only used a sub-region (350$\times$350 pixels) of these 
images to perform subtraction (see Figure 1). To minimize PSF
variation across the field, we varied the size of the sub-region 
using trial-and-error and found that a region of size 350$\times$350 
pixels (see Figure 2) gave the best image subtraction. While we set 
the background order to 2 ({\tt -bg 2} in {\tt HOTPANTS}) and the 
kernel order to 0 ({\tt -ko 0} in {\tt HOTPANTS}), the rest of the 
parameters were set to their default values. To be consistent, we 
always convolved with and normalized with respect to the template. 
The template--comprising of 31 images--was extracted by averaging 
all the aligned images with seeing less than 0.55$^{\prime\prime}$. 
The night of the 13$\textsuperscript{th}$ had the best seeing 
conditions, hence the template image was constructed from images 
obtained on this night only. We also tested the image subtraction 
analysis with a template constructed from 35 images with seeing 
better than 0.65$^{\prime\prime}$ from the night of the 
12$\textsuperscript{th}$ (a template independent of Feb 
13$\textsuperscript{th}$ data). The resulting ULX light curve was similar
within the error bars. A sample image before and after subtraction 
is shown in the top left and the top right panels of Figure 2, 
respectively.

We performed aperture photometry at 1000 randomly sampled locations
around the ULX (see Appendix A for aperture photometry methodology and
error estimation). We avoided regions close to saturated stars. By
choosing such a large number of points we ensured to sample the entire
region within plus or minus 100 pixels around the ULX. We then modeled
each of these relative background light curves with a constant model
which yielded a reduced $\chi^2$ between 0.5-1.5 providing confidence
in the image subtraction procedure. Seven of the sample background light curves
are shown in the bottom left panel of Figure 2.

Moreover, if the image subtraction was successful, the pixel values in
the image should be distributed symmetrically around some mean value. 
Figure 3 shows histograms of pixel values around X-1 (excluding saturated 
stars) from three different subtracted images. The histograms were 
constructed from a square region of width 100 pixels centered on the ULX. 
We excluded the ULX aperture of 6$\times$6 pixels centered on the ULX from 
these histogram calculations. It is clear that these histograms are 
symmetric and again validate the image subtraction procedure.

\subsubsection{NGC 5408 X-1's optical light curve:}
We first performed aperture photometry at the ULX's location 
in all the subtracted images. This gave us X-1's count rates 
(e$\textsuperscript{-}$/s) relative to the template image
($c_{i}$$\pm$$\delta{c_{i}}$, $i$=1...23). We then extracted 
X-1's count rate in the template used for image subtraction
($c_{template}$$\pm$$\delta{c_{template}}$). X-1's observed 
count rate was then estimated as the sum $c_{i}$ + 
$c_{template}$. The uncertainties on these rates were 
calculated as the sum in quadrature of $\delta{c_{i}}$ and 
$\delta{c_{template}}$. X-1's normalized optical light 
curve (fraction of its mean count rate) 
from the night of the 13$\textsuperscript{th}$ is shown 
in the bottom panel of Figure 4. Within the error bars, 
X-1's optical light curve is consistent with being 
constant giving a $\chi^2$ of 20 for 23 dof, when modeled
with a constant. 

Following the same procedure as described above (section 3.1.1 --
3.1.5) we constructed X-1's normalized optical light curves from the
nights of the 12$\textsuperscript{th}$ and the
14$\textsuperscript{th}$. These are shown in Figure 5 and they are
also consistent with being constant within the measured error
bars. The best-fit constant model yielded $\chi^2$/dof of 6/17 and
18/23 for the data on the 12$\textsuperscript{th}$ and
14$\textsuperscript{th}$, respectively. Note the larger error bar on
the 14$\textsuperscript{th}$ due to higher level of sky background.

On the night of the 12$\textsuperscript{th}$ we observed a
standard star field (NGC 2268) for $\approx$ 3 s in the 
g$\_$HIGH filter at an airmass of 1.02. Using a few bright,
unsaturated stars in this image, we estimated based on the STETSON catalog$\footnote{http://www.cadc-ccda.hia-iha.nrc-cnrc.gc.ca/en/community/STETSON/standards/}$ the photometric 
zero-point count rate of the CCD to be 
(8.68$\pm$0.36)$\times$10$^{10}$ e$\textsuperscript{-}$/s. 
The VLT/FORS2 g\_HIGH filter is close to the SDSS g$^{\prime}$ 
band. Using the filter transformations of Lupton (2005)$\footnote{https://www.sdss3.org/dr8/algorithms/sdssUBVRITransform.php\#Lupton2005}$ we 
estimated the mean V-band magnitude of NGC 5408 X-1's optical
counterpart on the 12$\textsuperscript{th}$, the 
13$\textsuperscript{th}$, and the 14$\textsuperscript{th}$ to
be 22.523$\pm$0.014, 22.539$\pm$0.015, and 22.563$\pm$0.016, 
respectively. Since the quoted error bars only take into 
account the statistical uncertainty, the true uncertainty 
in the zero-point--including the systematic uncertainty--is 
likely larger than this. Nevertheless, these 
values are consistent with prior studies 
of this object (Kaaret \& Corbel 2009; Gris{\'e} et al. 2012).
X-1's template count rate was estimated to be 
80.2$\pm$1.3 e$\textsuperscript{-}$/s. As a sanity 
check, we compared this value to the expected count 
rate using the VLT/FORS2 online exposure time calculator 
(ETC)$\footnote{https://www.eso.org/observing/etc/bin/gen/form?INS.NAME=FORS+INS.MODE=imaging. 
We used X-1's V-band mangiude of 22.4$\pm$0.6 as listed  
in Table 4 of Gladstone et al. (2013). Also for the ETC 
we assumed an airmass of 1.5 with a seeing of 
0.6$^{\prime\prime}$. The source spectrum was assumed to 
be a powerlaw with an index of -2 as dervied by Kaaret \& Corbel 
 (2009). Note that by default the ETC assumes an 
aperture radius equal to the seeing. We, however, use 
0.67$\times$seeeing.}$. The observed values are consistent 
with the values estimated by the ETC.

In order to investigate if X-1's optical counterpart
varied on a timescale of $\approx$ 24 hours, we first estimated 
its mean count rate on each night. We then divided each of those 
by the mean over three nights to extract a normalized count rate 
for each night. Similarly, we estimated the normalized count rate 
of a sample of seven nearby point sources. The RMS scatter in the 
normalized count rates of the field stars--on each night--was 
added as an additional uncertainty to X-1's normalized light curve 
measurement. The final long-term light curve of X-1 is shown in 
Figure 6. Since the evidence for variability on an $\approx$ day timescale 
is only significant at the 3$\sigma$ level, we consider it to be 
marginal.

\subsubsection{Intra-Night RMS variability Upper limit}
The amount of variability in a light curve can be quantified using the
so-called normalized excess variance ($\sigma_{NXS}^{2}$; Edelson et
al. 2002; Vaughan et al. 2003). The fractional root-mean-squared (RMS)
amplitude is simply
\[ 
F_{var} = \sqrt{\sigma_{NXS}^{2}} 
\] 
At the $n$-sigma level, a source is considered to be variable only if
$\sigma_{NXS}^{2}$ $>$ $n$$\times$err($\sigma_{NXS}^{2}$), where
err($\sigma_{NXS}^{2}$) is the error on the normalized excess
variance. Using the analytical expression for err($\sigma_{NXS}^{2}$)
derived by Vaughan et al. (2003), we estimate a 3-$\sigma$ upper limit
on X-1's fractional RMS optical variability on the 12$\textsuperscript{th}$, 
the 13$\textsuperscript{th}$, and the 14$\textsuperscript{th}$ to be 3.3\%, 
4.8\%, and 5.6\%, respectively.

\section{Analysis: X-ray Light Curve}
For this study, we only used data acquired on 2014 February 13 by the 
European photon imaging camera (EPIC; both {\tt pn} and {\tt MOS}) on 
board {\it XMM-Newton}. We used the latest standard analysis system 
({\tt XMMSAS}) version 14.0.0 to reduce the images, and extract the 
filtered EPIC event lists. All the event lists were screened with a 
standard filter of ({\tt PATTERN $<=$ 4}), to include only the single 
and the double pixel events, and events in the complete band pass of 
0.3-10.0 keV were considered for further analysis.

X-1's mean X-ray count rate was 1.288$\pm$0.006 counts s$^{-1}$.
To be able to compare with the optical light curve we show the 
normalized X-ray light curve from 2014 February 13 (simultaneous
with the optical observations) in the top panel of Figure 4. To test 
for variability we modeled the X-ray light curve with a constant. This
yielded a $\chi^{2}$ of 381 for 72 degrees of freedom (dof). This 
suggests that a constant model is strongly disfavored which is in 
agreement with earlier studies (e.g., Dheeraj \& Strohmayer 2012;
Caballero-Garc{\'{\i}}a et al. 2013, etc). To quantify the variability, we calculated the normalized excess variance 
($\sigma_{NXS}^{2}$) and the fractional root-mean-squared (RMS)
amplitude following Vaughan et al. (2003). Using a light curve bin 
size of 440 s as in the optical light curve, we estimate the
fractional RMS amplitude in X-rays to be 9.0$\pm$0.5\%.

\section{X-ray -- Optical Cross-Correlation}
To search for a reverberation signal we evaluated the discrete cross
correlation function (DCF; Edelson et al. 1988) between the X-ray and
the optical light curves taken on the 13$\textsuperscript{th}$ (see
the left panel of Figure 7). There is no evidence for a statistically
significant correlation between the X-ray and the optical variability.

We estimated the DCF's error bars and the significance contours
following these steps:

1) First, we constructed $N_{sample}$ = 10$^{6}$ synthetic X-ray and
optical light curves. These were drawn from the measured light
curves--assuming for each time stamp in the measured light curve--the
count rates are Gaussian distributed with a mean equal to the measured
count rate and a standard deviation, $\sigma$, equal to the error bar.

2) We then extracted $N_{sample}$ corresponding DCFs using 
the same input parameters as the observed DCF. The error bar 
at a given lag was then estimated to be twice the standard 
deviation in the DCF measurement within these $N_{sample}$ values.
In essence, an error bar at a given lag bin represents 
4$\sigma$ deviations. 

3) Finally, we estimated the 3 and the 4$\sigma$ confidence 
contours on the DCF using model-independent Monte Carlo 
simulations as follows:

$\bullet$ For each of the synthetic optical light 
curves derived in step 1, we randomized the count rate 
values but retained the time stamps.

$\bullet$ We then calculated the DCF between these randomized 
synthetic optical light curves and a synthetic X-ray light curve.
The time stamps of the synthetic X-ray light curve were kept
the same (null hypothesis).

$\bullet$ After repeating the above two steps $N_{sample}$ number
of times we estimated--for each lag--3$\sigma$ and 
4$\sigma$ significance levels.

It can be seen from Figure 4 that both the X-ray and the optical light
curves are regularly sampled for the most part.  There are only a few
segments of the light curves where data was unevenly sampled. After
interpolating at these handful of data points, we constructed the
regular cross correlation function (CCF) using the {\tt IDL} function
{\tt C\_CORRELATE}.  The corresponding CCF is shown in the right
panel of Figure 7. We set the lag limits to ensure that the entire
optical light curve is within the bounds of the X-ray light
curve. Thus, the maximum and the minimum lags were set to 15000 s and
-7500 s, respectively. In this scheme, a positive lag would imply
that the optical lags the X-ray and {\it vice versa for a negative
  lag}.  As expected, the regular CCF (right panel of Figure 7) is
very similar to the DCF (left panel of Figure 7). The error bars and
the confidence contours were extracted using the same
model-independent Monte Carlo methodology as described for the DCF.
Again, the CCF indicates that there are no statistically significant
features in the CCF.

\section{Discussion}

Timing studies of stellar-mass black hole binaries using simultaneous
X-ray and optical/UV monitoring have found two main types of correlated
behavior: (1) straightforward reprocessing lags where the optical
emission is correlated with the X-ray but delayed by a time-scale
consistent with the light travel time to reprocessing sites within the
binary system, and (2) complex, non-reprocessing time lags where the
optical emission can lead the X-rays (see, Gandhi et al. 2008; Durant
et al.  2008; Motch et al. 1983, 1985 and references therein). Indeed,
it appears that both processes can be present together with the
non-reprocessing lags becoming dominant at time-scales much shorter
than the expected light travel time delays (Gandhi et al.
2010). Further, evidence suggests that synchrotron emission from the
jet may be responsible for this fastest, correlated variability seen
to date in black hole systems (Hynes et al. 2003; Kanbach et
al. 2001). For the purposes of mass estimates from reverberation
mapping one strictly requires the first type of correlated
variability.

The most clear-cut examples of such reprocessing are provided by
simultaneous optical/X-ray observations of thermonuclear X-ray bursts
from accreting neutron stars (see McClintock et al. 1979; Pedersen et
al. 1982; Hynes et al. 2006). However, reprocessing time-scales
(optical lagging the X-ray) have now also been measured from a handful
of stellar-mass black holes including GX 339-4 (Gandhi et al. 2010),
GRO J1655-40 (Hynes et al. 1998; O'Brien et al. 2002), XTE J1118+480
(Hynes et al. 2003), and {\it Swift} J1753.5-0127 (Hynes et al. 2009). The time
lags measured in these cases ($\approx$ 5 - 20 s) are broadly
consistent with their binary separations and expected accretion disk
sizes. In the case of GRO J1655-40 the inferred lag of $\approx 20$ s
(O'Brien et al. 2002) can be combined with the measured widths of
double-peaked HeII (4686 $\textrm {\AA}$) lines arising in the accretion
disk (Soria et al. 2000) to arrive at a mass estimate based on
Equation (1).  Taking half the peak separation of $\approx 550$ km
s$^{-1}$ for the HeII lines measured in 1996 June by Soria et
al. (2000) as representative of velocities in the disk, one obtains a
mass estimate of $M=3.4 M_{\odot}$ that is consistent with the mass
measurements based on the radial velocity curve for GRO J1655-40 (Beer
\& Podsiadlowski 2002; Greene et al. 2001).  This provides a basic
consistency check that the method can provide reliable mass estimates.

On timescales relevant to reverberation, black hole binaries
typically show optical fractional variability levels (RMS) in the range from
$\approx 3 - 10 \%$ (Hynes et al. 1998; Gandhi 2009). Assuming X-1
has a mass greater than $500 M_{\odot}$ (Strohmayer \&
Mushotzky 2009; Cseh et al. 2013), the corresponding range of lag
time-scales expected is of the order of 1000 s. We did not detect
statistically significant variability from X-1 on such time-scales,
and our $3\sigma$ upper limit of 3.3\% is roughly consistent with the lower
half of the optical RMS amplitude range quoted above. If X-1 were varying in the optical
closer to the high end of this range then we would have likely
detected such variability. Evidence of absorption-like dips in the
X-ray flux from X-1 is suggestive of a high inclination (Pasham \&
Strohmayer 2013; Gris{\'e} et al. 2013), and it is conceivable that a high
inclination could reduce the reprocessing response compared to a lower
inclination system (see, for example, Figure 3 of O'Brien et al. 2002).
Moreover, during reprocessing in X-ray binaries, the typical value of 
the ratio of the fractional RMS variability amplitude of the X-rays 
to the optical is roughly 5 (e.g., see Hynes et al. 1998; O'Brien et 
al. 2002). Assuming a similar ratio for X-1, the expected RMS optical 
variability is roughly 2\% (9/5). This is still a factor of two
lower than the fractional RMS upper limit we reached with the VLT data on the 
13$\textsuperscript{th}$ (4.8\%).

While this first attempt at measuring an optical/X-ray reverberation
lag in X-1 was unsuccessful, it has ``bounded'' the problem and
demonstrated what is required for success. Our best limits were
achieved on the night of 12$\textsuperscript{th}$ February and 
corresponded to times of lowest sky background (largest moon angles). 
These observations are particularly challenging because of the need to
simultaneously satisfy several observing constraints. First, the
target must be visible from the optical telescope at low to modest
airmass values, while also being simultaneously observable with
{\it XMM-Newton}. Second, for optimum sensitivity the moon should also be
below the horizon to limit the sky background. For these observations
we achieved the first constraint, but not the latter.  Moreover,
assuming the lag time-scale is $\sim$1000 s, then significantly more
simultaneous X-ray/optical data is desirable than over a single
night. It should be possible to achieve sensitivity to variability at
the few percent level if all the above constraints can be satisfied.
Nevertheless, even if the observing constraints are satisfied,
ground-based optical observations remain at the mercy of weather and
seeing conditions, and for this reason Hubble Space Telescope (HST)
observations may provide a better solution.

Finally, we reiterate that the time resolution and the overall temporal 
baseline of the optical and the X-ray light curves set the lower and the 
upper limits on the black hole masses that can be probed with the 
reverberation technique proposed here. With the current optical 
telescopes, ULX optical counterparts require exposures on the order of 
hundred seconds to reach the signal-to-noise to detect a reverberation signal. Thus
with current technology, this method is more sensitive to IMBHs than 
stellar-mass black holes. However, the mass range can be extended to 
stellar-mass black holes with the next generation telescopes such as 
the James Webb Space Telescope (JWST) and the Thirty Meter Telescope (TMT).

{\bf Acknowledgements} We would like to acknowledge the excellent
support of the ESO staff at Paranal and particularly the FORS2
instrument scientist (Dr. Henri Boffin) without whom we could not have
obtained the optical observations. DRP would like to thank Sylvain Veilleux
and Rob Olling for discussions about optical photometry.

\begin{table}
    \caption{{ Setup for VLT/FORS2 observations of NGC 5408 X-1}}\label{Table1} 
{\small
\begin{center}
   \begin{tabular}[t]{lc}
	\hline\hline \\
{\bf CCDs}		& MIT/LL Mosaic \\
			& CCID20-14-5-3 \\
\\
{\bf Number of}		& 2 chips of \\
{\bf pixels}		& 2048$\times$1034 each\tablenotemark{1, 2} \\ 
\\
{\bf Gain}		& 0.8 e$\textsuperscript{-}$ ADU$\textsuperscript{-1}$ \\
\\
{\bf Filter}		& g\_HIGH \\
\\
{\bf Pixel Size}	& 0.252$^{\prime\prime}$$\times$0.252$^{\prime\prime}$\tablenotemark{1} \\
\\
{\bf Exposure}\tablenotemark{3}		& $\approx$ 120 s/165 s \\
\\
{\bf Readout Noise}	& 2.7 e$\textsuperscript{-}$  pixel$\textsuperscript{-1}$  \\
\\
    \hline\hline
    \end{tabular}
\end{center}
}

\tablenotemark{1}{After the standard 2$\times$2 spatial binning.}\\
\tablenotemark{2}{X-1 was always situated on Chip 1. Chip 2 data was not used in this work.}\\
\tablenotemark{3}{Mean exposure time per individual image. The exposures were each 120 s on 
the nights on the 13$\textsuperscript{th}$ and the 14$\textsuperscript{th}$, while 165 s 
exposures were taken on the night of the 12$\textsuperscript{th}$.}\\
\end{table}


\begin{figure}[ht]

\begin{center}
\includegraphics[width=6.5in, height=6.25in, angle=0]{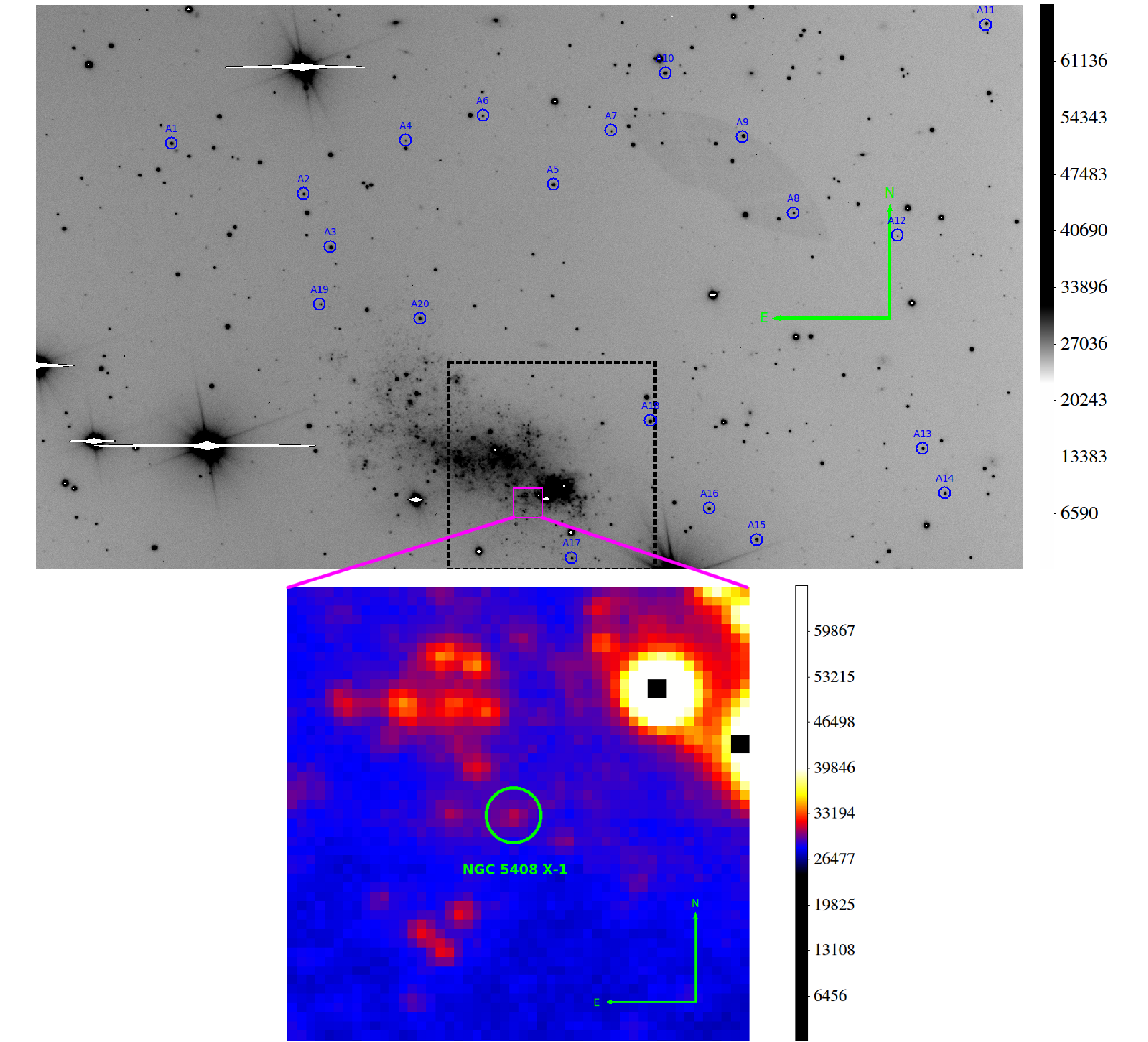}
\end{center}
\vspace{-.35cm} 
{\textbf{Figure 1:} {\it Top Panel:} Sample VLT/FORS2
image of NGC 5408 X-1 showing the full field of view. The sources used
to check image alignment are indicated by blue circles (also see Table
2). The dashed black box is the sub-region used for image subtraction
with {\tt HOTPANTS} (see Figure 2). {\it Bottom Panel:} Zoomed-in view
of the region around the ULX NGC 5408 X-1 (green circle). The north
and the east arrows are each 50$^{\prime\prime}$, and
2.5$^{\prime\prime}$ in the top and the bottom panels, respectively.}
\label{fig:figure1}
\end{figure}



\begin{figure}[ht]

\begin{center}
\includegraphics[width=6.25in, height=5.85in, angle=0]{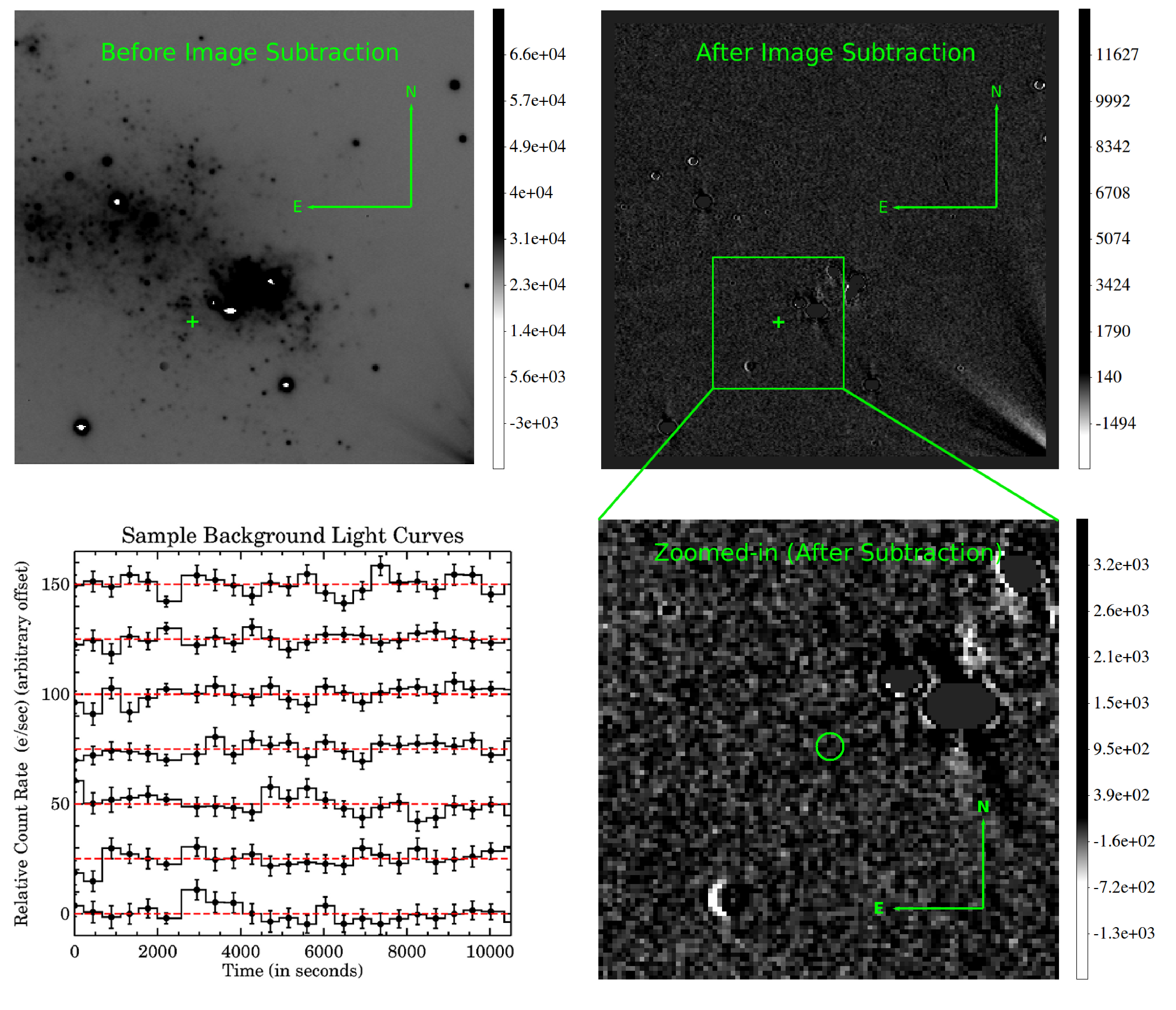}
\end{center}
\vspace{-.35cm} {\textbf{Figure 2:} A sample FORS2 image before (top
left panel) and after (top right panel) image subtraction. The arrows
pointing north and east are each 20$^{\prime\prime}$ long. In both the
images the ULX's location is marked with a green cross. {\it Bottom
  Right}: A zoomed-in image of 100$\times$100 pixels wide centered on
X-1. The north and the east arrows are each 5$^{\prime\prime}$ long
and the circle indicating the ULX's position has a radius of 3 pixels
(the aperture radius used for photometry). {\it Bottom Left:} Sample
relative background light curves from the night of the
13$\textsuperscript{th}$ to validate image subtraction. They are all
consistent with being constant.  }
\label{fig:figure1}
\end{figure}



\begin{figure}[ht]

\begin{center}
\includegraphics[width=6.5in, height=2.1in, angle=0]{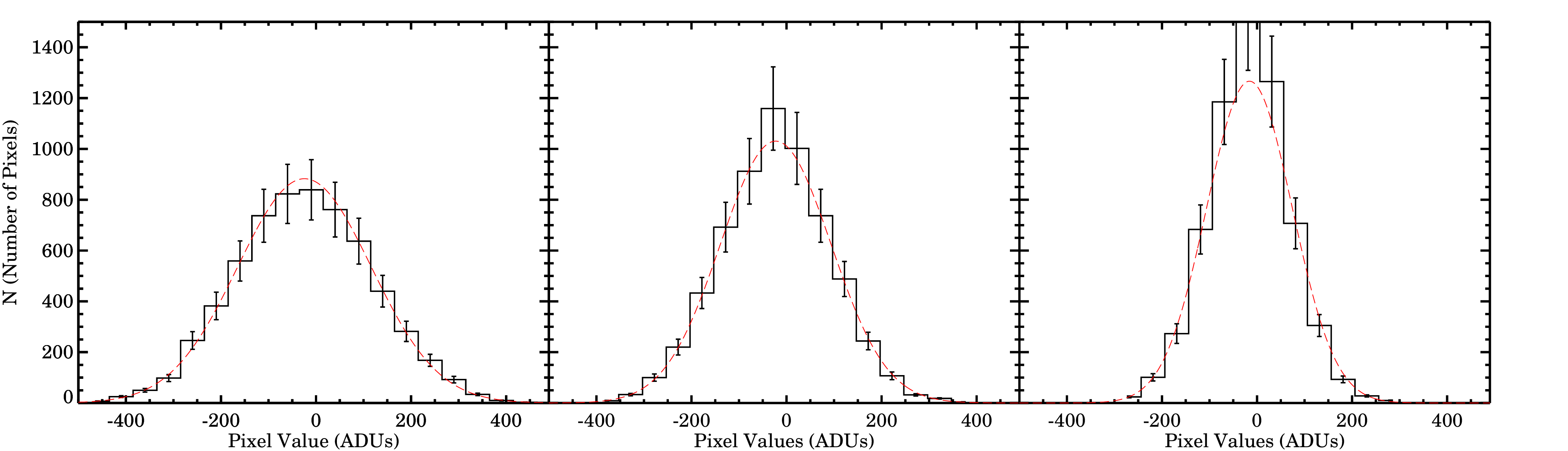}
\end{center}
\vspace{-.35cm} {\textbf{Figure 3:} 
Histograms of pixel values around
X-1 ($\pm$50 pixels centered on X-1) in three subtracted
images. Clearly, they are all symmetrically distributed. These are 
reasonably represented by a Gaussian model (red curves). Deviations are likely due
to systematic uncertainties introduced by the image subtraction process (mis-alignment, PSF mismatch, etc).  }
\label{fig:figure1}
\end{figure}



\begin{figure}[ht]

\begin{center}
\includegraphics[width=6.5in, height=4.33in, angle=0]{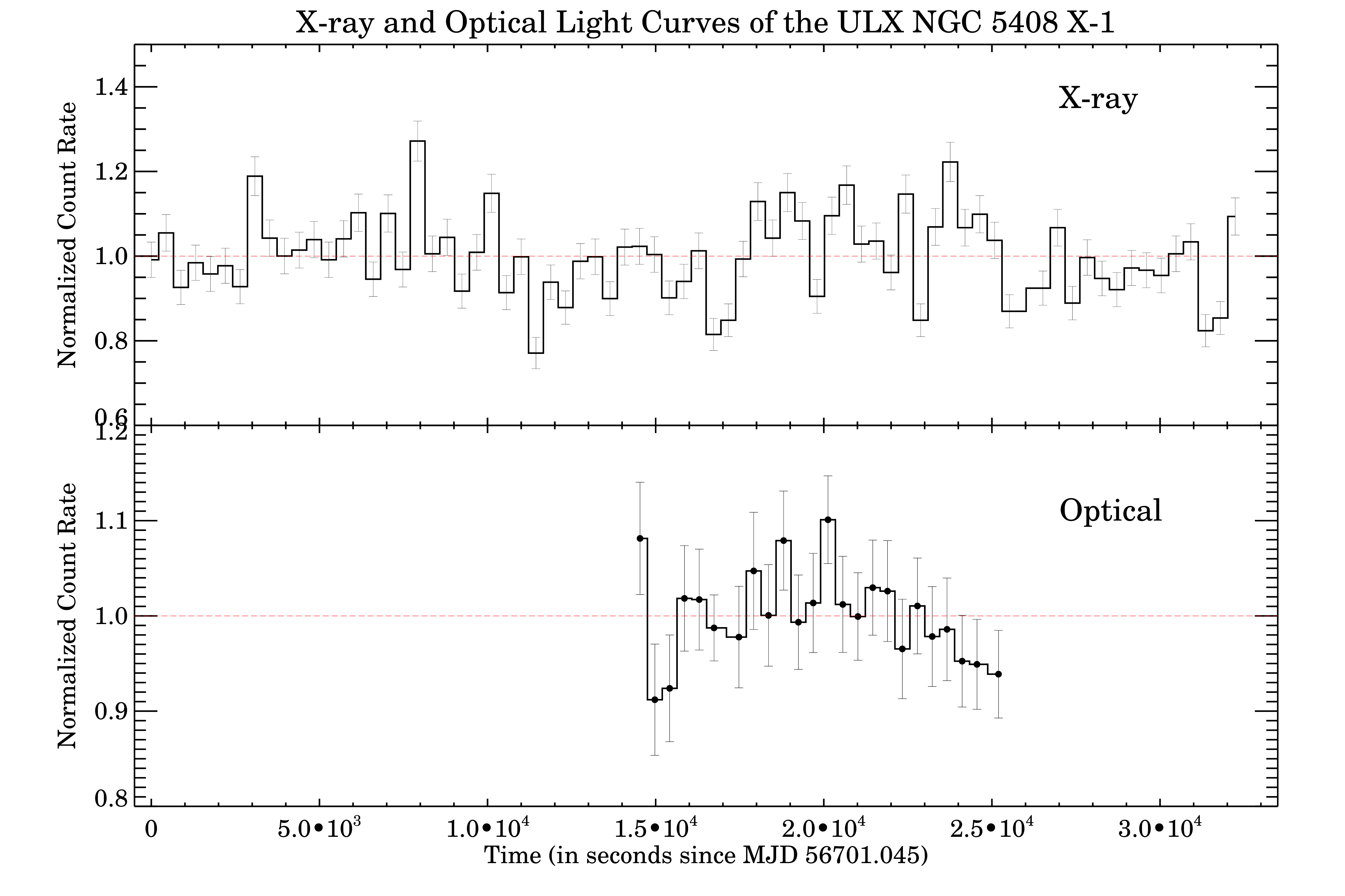}
\end{center}
\vspace{-.35cm} {\textbf{Figure 4:} 
Normalized X-ray (top) and
optical (bottom) light curves of NGC 5408 X-1 taken on 2014 February
13. While the X-rays are variable (fractional RMS variability of
9$\pm$0.5\%), the optical light curve--within the error bars--is
consistent with being constant. An upper limit on the fractional RMS
amplitude in order to be considered variable at the 3$\sigma$ level is
4.8\%.  For consistency, both the X-ray and the optical light curves
have the same time bin size of 440 s.  }
\label{fig:figure1}
\end{figure}



\begin{figure}[ht]

\begin{center}
\includegraphics[width=5in, height=6.85in, angle=0]{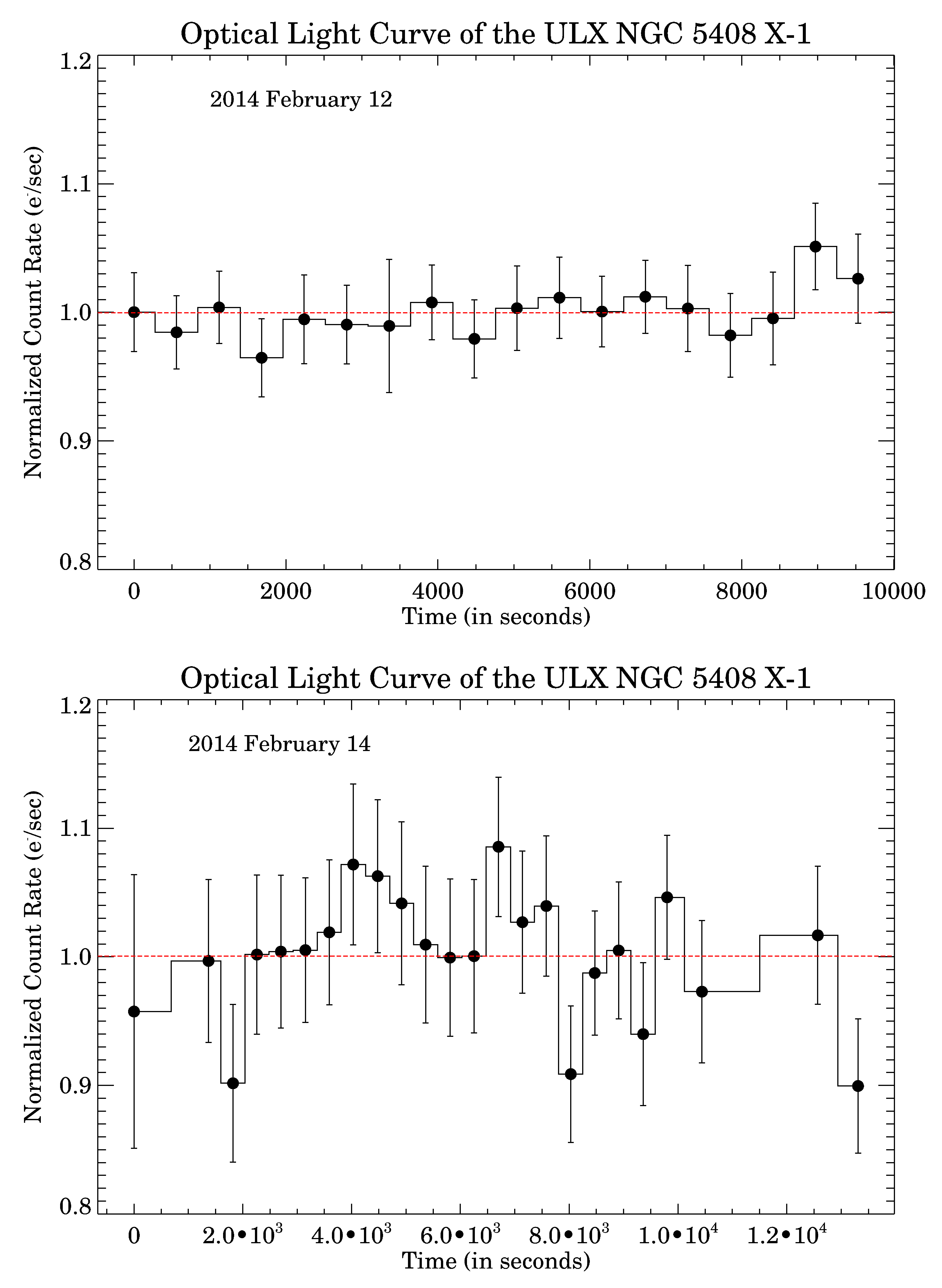}
\end{center}
{\textbf{Figure 5:} X-1's normalized optical light 
curve from 2014 February 12 (top) and 2014 February 
14 (bottom). Similar to the optical light curve from
2014 February 13, these are also consistent with 
being constant. The 3$\sigma$ upper limits on the optical 
variability are 3.3\% and 5.6\% for the 
12$\textsuperscript{th}$ and the 14$\textsuperscript{th}$,
respectively.
}
\label{fig:figure1}
\end{figure}



\begin{figure}[ht]

\begin{center}
\includegraphics[width=5.35in, height=4.in, angle=0]{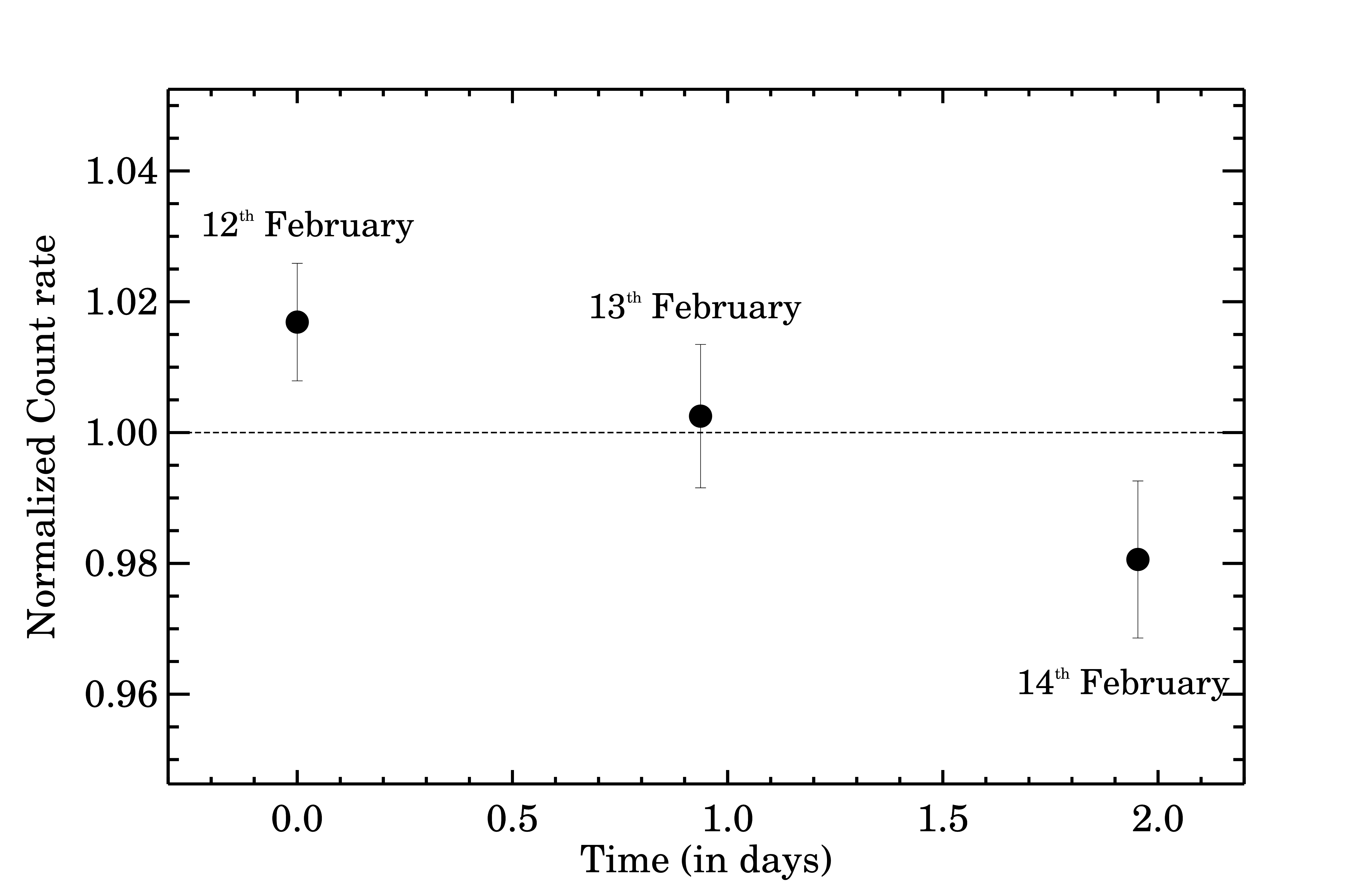}
\end{center}
{\textbf{Figure 6:} NGC 5408 X-1's normalized optical 
light curve over the three nights in 2014. The evidence for 
variability on a day timescale is only marginal 
($\approx$ 3$\sigma$).
}
\label{fig:figure1}
\end{figure}



\begin{figure}[ht]

\begin{center}
\includegraphics[width=6.35in, height=3.in, angle=0]{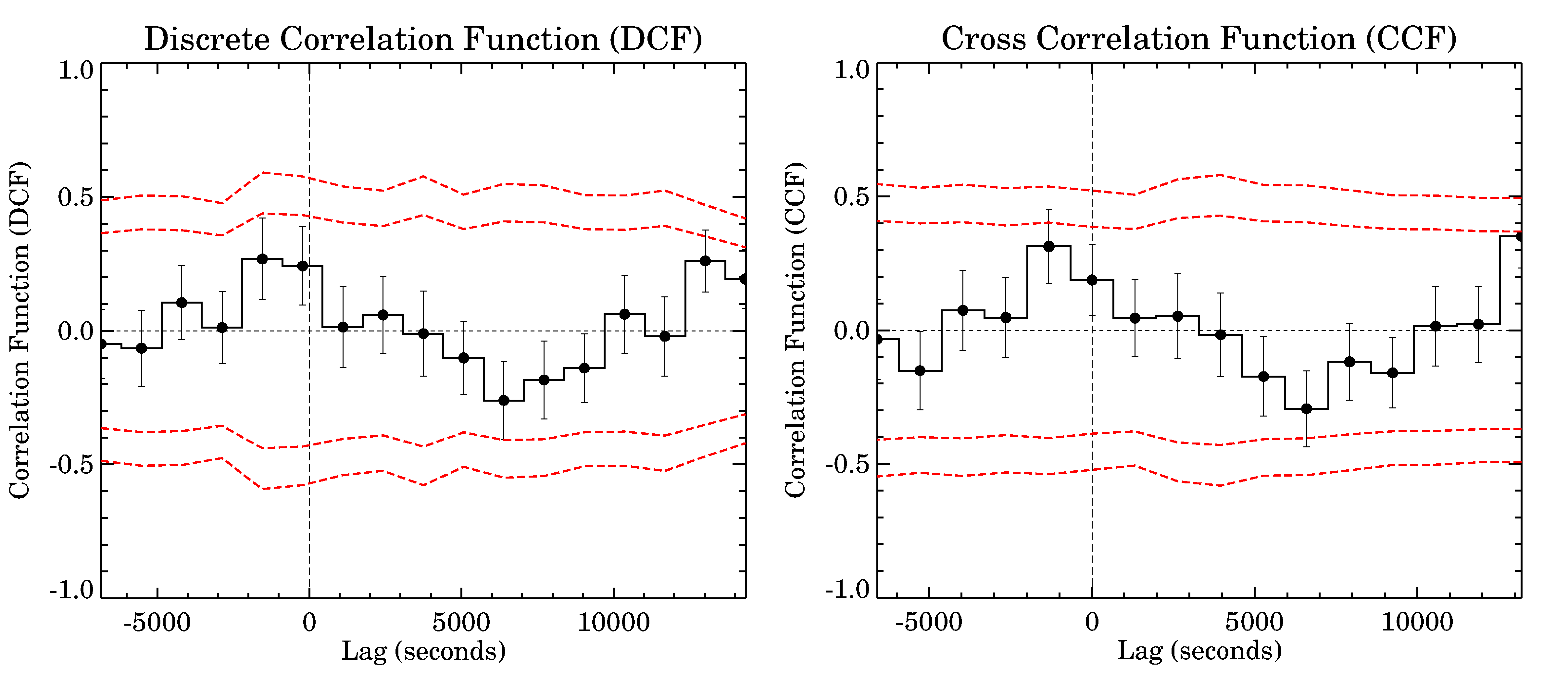}
\end{center}
{\textbf{Figure 7:} Discrete Cross Correlation 
Function (DCF) (left panel) and a regular cross 
correlation function (CCF) (right panel) between
the X-ray and the optical light curves of NGC 5408 
X-1. Clearly, they both are consistent with each
other. The 3 and the 4$\sigma$ significance contours 
(dashed red curves) were calculated using 
model-independent Monte Carlo simulations (see 
section 5). There are no statistically significant 
features. A positive lag in these plots would imply 
that the optical lags the X-ray emission.   
}
\label{fig:figure1}
\end{figure}


\begin{table}
    \caption{{ Summary of Image Alignment for FORS2 data taken on 2014 February 13.}}\label{Table2} 
{\small
\begin{center}
   \begin{tabular}[t]{cccc}
	\hline\hline \\
Source ID\tablenotemark{1}	&		$\alpha$			&	$\delta$	&	RMS\tablenotemark{2} \\
		&	(right ascension)			& (Declination)		&	(\% of a pixel)	\\
\hline\hline \\

A1 &	14:03:33.16 &	-41:20:24.7 &	7.2\\
A2 &	14:03:28.15 &	-41:20:46.4 &	6.4\\
A3 &	14:03:27.14 &	-41:21:09.0 &	5.9\\
A4 &	14:03:24.29 &	-41:20:23.7 &	8.8\\
A5 &	14:03:18.69 &	-41:20:42.5 &	5.7\\
A6 &	14:03:21.36 &	-41:20:13.0 &	7.8\\
A7 &	14:03:16.49 &	-41:20:19.5 &	7.1\\
A8 &	14:03:09.58 &	-41:20:54.7 &	9.6\\
A9 &	14:03:11.52 &	-41:20:22.3 &	3.5\\
A10 &	14:03:14.44 &	-41:19:54.8 &	4.3\\
A11 &	14:03:02.30 &	-41:19:34.4 &	8.7\\
A12 &	14:03:05.65 &	-41:21:04.2 &	11.8\\
A13 &	14:03:04.68 &	-41:22:35.2 &	10.5\\
A14 &	14:03:03.83 &	-41:22:54.2 &	6.7\\
A15 &	14:03:10.98 &	-41:23:14.1 &	8.3\\
A16 &	14:03:12.77 &	-41:23:00.7 &	8.6\\
A17 &	14:03:18.00 &	-41:23:21.8 &	10.5\\
A18 &	14:03:15.01 &	-41:22:23.4 &	5.7\\
A19 &	14:03:27.56 &	-41:21:33.6 &	8.2\\
A20 &	14:03:23.75 &	-41:21:39.7 &	5.9\\
\\
    \hline\hline
    \end{tabular}
\end{center}
}

\tablenotemark{1}{See Figure 1 for the location of these sources.}\\
\tablenotemark{2}{The root mean squared deviation of the centroid of the source as estimated over all the 73 images. It is expressed as percentage of a pixel.}\\
\end{table}

\newpage
\section*{Appendix} 
\begin{center} {\bf Methodology for Obtaining the Optical Count rate and their Error Bars}\end{center}
We performed aperture photometry at a given location on the 
subtracted images as follows.

(1) First, we estimated the total number of counts in a 
circular aperture of radius 3 pixels ($N_{src,uncorr}$). 
For a Gaussian point spread function (PSF), it can be 
shown straightforwardly that the signal-to-noise ratio 
is maximum when the aperture radius is $\approx$ 
0.68$\times$(Full-Width-Half-Maximum). In our case, the 
Full Width Half Maximum (FWHM) is equal to the seeing. 
Now, considering the worse seeing of our images of 
1.0$^{\prime\prime}$ or $\approx$ 4 pixels,
this corresponds to 0.68$\times$4 $\approx$ 3 pixels.

(2) We then calculated the median background counts per
pixel using a nearby region. This value was multiplied 
by the total number of background pixels, and then by 
the ratio of the source to the background aperture 
($\eta$) to obtain the total background counts corrected 
for the difference between the source and the background 
area ratio. We call this quantity $N_{bkg,corr}$.

(3) The background-corrected source counts were then 
calculated as 

\[ N_{src,corr} = N_{src,uncorr} - N_{bkg,corr} \]


\noindent The error on $N_{src,corr}$ was estimated by a simple 
propagation of uncertainty in quadrature as,

\[
\sigma_{N_{src,corr}} = \sqrt{ \sigma^{2}_{N_{src,uncorr}} + \sigma^{2}_{N_{bkg,corr}} }
\]

\noindent where, 
\[
\sigma^{2}_{N_{src,uncorr}} = |N_{src,uncorr}|
\]

\[
\sigma^{2}_{N_{bkg,corr}} = {\rm RMS}_{bkg}\times\sqrt{N_{bkg,pix}}\times\sqrt{\eta} 
\]

\noindent $N_{bkg,pix}$ is the total number of background pixels, and RMS$_{bkg}$
is root-mean-squared variance in the background pixel 
values calculated as
\[
\textrm{RMS}_{bkg} = \sqrt{\sum\frac{(z-z_{mean})^2}{N_{bkg,pix}}}
\]

$z$ are the pixel values and $z_{mean}$ is the mean of 
those values. 
\vfill\eject

\vfill\eject

\end{document}